\documentclass[journal,onecolumn]{IEEEtran}
\IEEEoverridecommandlockouts
\usepackage{cite}
\usepackage[T1]{fontenc}
\usepackage{amsmath}
\usepackage{bm}
\usepackage{amssymb}
\usepackage{amsfonts}
\usepackage{algorithmic}
\usepackage{textcomp}
\usepackage{xcolor}
\usepackage[utf8]{inputenc}
\usepackage{mathtools}
\usepackage{graphicx}
\usepackage{subcaption}
\usepackage{booktabs}
\usepackage{multirow}
\usepackage{caption}
\usepackage{arydshln}

\def\BibTeX{{\rm B\kern-.05em{\sc i\kern-.025em b}\kern-.08em
    T\kern-.1667em\lower.7ex\hbox{E}\kern-.125emX}}
\begin{document}

\title{Impact of die carrier on reliability of power LEDs\\

}

\author{\IEEEauthorblockN{Shusmitha Kyatam} \\
\IEEEauthorblockA{\textit{Instituto de Telecomunicações e Departamento de Electrónica, Telecomunicações e Informática,} 
\textit{University of Aveiro,}
3810-193 Aveiro, Portugal \\
shusmithakyatam@gmail.com} \\
\and
\IEEEauthorblockN{Luis N. Alves} \\
\IEEEauthorblockA{\textit{Instituto de Telecomunicações e Departamento de Electrónica, Telecomunicações e Informática,}
\textit{University of Aveiro,}
3810-193 Aveiro, Portugal \\nero@av.it.pt} \\
\and
\IEEEauthorblockN{Stanislav Maslovski} \\
\IEEEauthorblockA{\textit{Instituto de Telecomunicações e Departamento de Electrónica, Telecomunicações e Informática,}
\textit{University of Aveiro,}
3810-193 Aveiro, Portugal \\stas@av.it.pt} \\
\and
\IEEEauthorblockN{Joana C. Mendes} \\
\IEEEauthorblockA{\textit{Instituto de Telecomunicações e Departamento de Electrónica, Telecomunicações e Informática,} 
\textit{Campus de Santiago,}
3810-193 Aveiro, Portugal \\
joanacatarina.mendes@ua.pt}
}

\maketitle

\begin{abstract}
High power light emitting diodes (LEDs) suffer from heating effects that have a detrimental impact on the devices characteristics. The use LED carriers with high thermal conductivity promotes the extraction of heat away from the LED junction. Different materials can be used for this purpose, such as alumina, aluminium nitride, and silicon. Diamond has also been gaining momentum for demanding heat management applications. 
In order to evaluate the impact of the different carriers on the reliability of the devices, the junction temperature of Cree\textregistered\space white Xamp\textregistered\space XB-D LEDs was obtained with Ansys for various carrier at different LED current levels. The impact of the junction temperature on the LED’s lifetime, emission intensity, footprint and wavelength stability was then evaluated for each carrier based on the datasheet of the devices. The results provide additional knowledge regarding the impact of the carrier on the performance of the LED.
\end{abstract}


\section{Introduction}\label{sec:introduction}
High power light emitting diodes (LEDs) have revolutionized lighting applications. LEDs are compact devices with high lighting capability and spend only a fraction of the energy of filament bulbs. Nevertheless, these devices suffer from self-heating issues. Non-radiative recombination in the LED active region generates most of the heat at low current levels and, at high current levels, the parasitic resistances of the contacts and cladding layers provide an additional source of heat. Considering all these contributions, the efficiency of the LEDs stays below 30\%. As a rule of thumb, LED lumen output typically decreases 0.3-0.5\% for each 1\textdegree C increase in the junction temperature ($T_{\rm J}$)~\cite{Liu2019}. Similarly to what happens with other semiconductor devices, the increase of the $T_{\rm J}$ also has a negative impact in both the life time and reliability of the LEDs. Finally, the energy gap of the semiconductor depends on the temperature~\cite{Schubert2006}; as a consequence, the wavelength of the emitted light increases with $T_{\rm J}$. While the stability of the wavelength alone is not relevant for lighting applications, the associated change in the chromaticity of the LEDs may induce changes in the perception of object colors.

In order to facilitate the extraction of light from the device, the LED die is typically encapsulated inside a dome-shaped material with a large refractive index; however, dome materials are typically poor heat conductors, and this hinders the removal of heat by convection. To promote the efficient removal of the heat from the die, researchers and manufacturers have proposed different solutions that minimize the thermal resistance of the package. As an example, the integration of a copper (Cu) heat spreader increases the power efficiency by $3$\% when compared to a conventional package~\cite{Horng2009}. The use of flip-chip architecture provides an efficient means to decrease the thermal resistance of the package, especially if combined with carriers with large thermal conductivity ($\kappa$). Among the typically used materials one can find alumina (Al\textsubscript{2}O\textsubscript{3})~\cite{Liang2017}, silicon (Si)~\cite{Tsai2014} or aluminium nitride (AlN)~\cite{Shatalov2005a,Tang2010a}. Chun \textit{et al.}~\cite{Chun2007} went one step beyond and integrated the LED die with a Si thermoelectric cooler (TEC) MEMS using flip-chip. The impact of the solder pads, solder bumps and die attach~\cite{Tsai2014,Fan2008,Liu2019,Liang2017}, and underfills~\cite{Tsai2014, Tang2010a} has also been evaluated by different researchers. 

The carrier has an obvious impact on the thermal resistance of a LED package. In the case of surface mount devices (SMD), for instance, if one neglects the die attach and solder layers, the heat generated in the LED die active layers has to travel through the die substrate and carrier (when die and carrier are assembled via wire bonding) or through the carrier alone (in case of flip-chip architecture) before it reaches the back side of the LED package. Thus, for a given package layout, $T_{\rm J}$ depends on the carrier and materials with high $\kappa$ are desired.

Carbon-based materials, such as diamond, highly oriented pyrolytic graphite (HOPG), and graphene, are obvious candidates. Due to its structure, HOPG is an anysotropic thermal conductor ($\kappa_{\rm in-plane}\simeq$2000~W/(m$\cdot$K) while $\kappa_{\rm out-plane}$ is two orders of magnitude lower~\cite{Balandin2011}). If assembled with the die adequately, HOPG carriers could facilitate the transfer of the heat generated in the die towards the back side of the carrier. However, HOPG is also electrically conductive, and this creates some problems with respect to the insulation of the terminals of the p-n junction. Graphene features an even higher in-plane $\kappa$ (2000-4000~W/(m$\cdot$K)), however graphene sheets are also electrically conductive. In addition, since graphene is a 2D material (monolayer thickness $\simeq$3.35~\AA~\cite{Pop2012}), one can't take advantage of the in-plane $\kappa$ and promote the transfer of heat to the back of the device. Despite these limitations, one can find recent reports describing the use of graphene for thermal management applications~\cite{Huang2021}. 

Diamond is an isotropic material that features simultaneously a high thermal conductivity (2200~W/(m$\cdot$K), increasing to 3300~W/(m$\cdot$K) in the case of isotopically pure material) while being an electric insulator, with a breakdown field as high as 2$\times$10\textsuperscript{7}~V/cm~\cite{Gracio2010}. Given to these extreme properties, diamond has been used for the thermal management of electronic components at various levels. Its use as a heat sink material was proposed back in 1967, when Swan reported that Si avalanche diodes mounted on a single crystal diamond carrier achieved a continuous power density more than twice the one obtained with copper heat sinks~\cite{Swan1967}. Other reports on the use of diamond as a heat sink followed, a complete list can be found in~\cite{Neto_HS_2019}. Free standing diamond has also been used as a sub-mount for lasers~\cite{Kemp2013,Rotter2009} as well as a circuit board integrated with water-cooling channels~\cite{Apollo2018}. Gallium nitride (GaN) high electron mobility transistors (HEMTs) also benefit from the integration with diamond films~\cite{Jia2019,Cheng2020,Ahmed2020,Smith2020,Helou2020}.

Diamond has also been used to improve the thermal management of power LEDs at different levels. Horng \textit{et al.}~\cite{Horng2009} mounted the LED die on a diamond-coated copper heat sink and obtained an 11\textdegree C reduction in $T_{\rm J}$ in comparison to mounting the LED on a conventional MCPCB. Chen \textit{et al.}~\cite{Chen2008a} reported a 20\textdegree C reduction in $T_{\rm J}$ at 1~A when the LED chips were bonded on Si substrates coated with 20~$\mu$m of diamond. Fan \textit{et al.}~\cite{Fan2011} replaced the Cu heat sink of white power LEDs with a diamond/Cu composite material and obtained a reduction on the weight of the internal heat sink by more than 35\% and simultaneously a decrease in the thermal resistance and $T_{\rm J}$ by as much as 10.5\% and 33.3\%, respectively. Diamond films have also been integrated directly with the LED die in grooves etched on the upper ITO layer of LEDs~\cite{Xie2020}.

Despite the promising results of these different approaches, the use of a diamond plate as the carrier of power LED dies has not, to the authors' best knowledge, been reported yet. This paper aims at estimating the expected benefits of such an approach. To this end, the temperature profile of Cree\textregistered\space white XLamp\textregistered\space XB-D power LEDs were obtained with Ansys. Different carriers were considered in the simulations: AlN (the actual carrier, as described by the manufacturer), Al\textsubscript{2}O\textsubscript{3}, Si, and diamond. The results were used to estimate the acceleration factor (AF) and the change in the relative luminous flux (RLF) when the AlN carrier is replaced with a material with different thermal properties. The dependency of the wavelength on the current level of monochromatic LEDs mounted on different carriers was also estimated. 

\section{Simulations and Modeling}\label{sec:experimental}

\subsection{LED structure}\label{LED_and_boards}
\begin{figure}
	\centering
	\begin{subfigure}[b]{0.25\textwidth}
		\includegraphics[width=\textwidth]{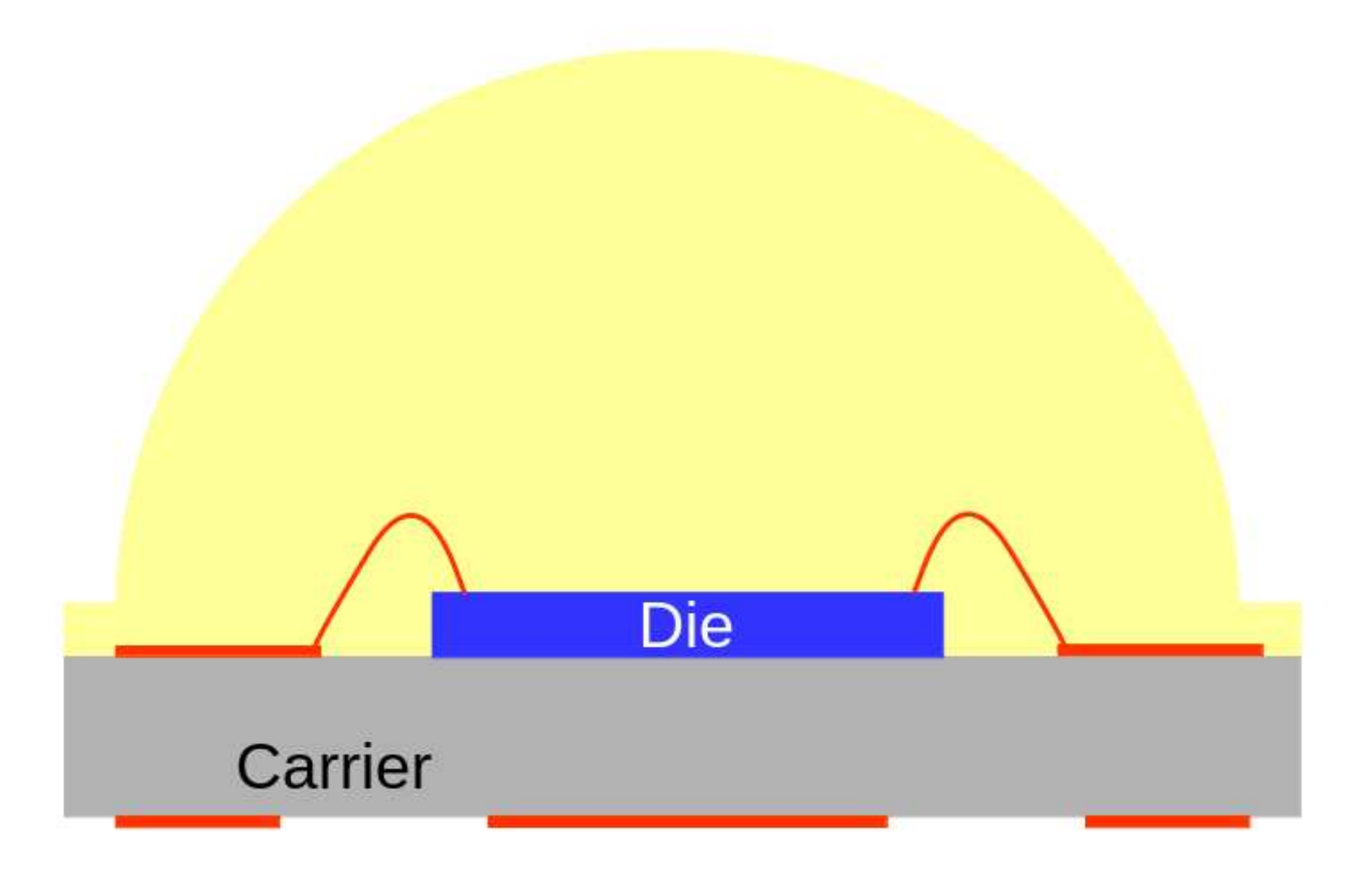}
		\caption{}
		\label{fig:LED_structure_CSV}
	\end{subfigure}
	\qquad
	~ 
	\begin{subfigure}[b]{0.15\textwidth}
		\includegraphics[width=\textwidth]{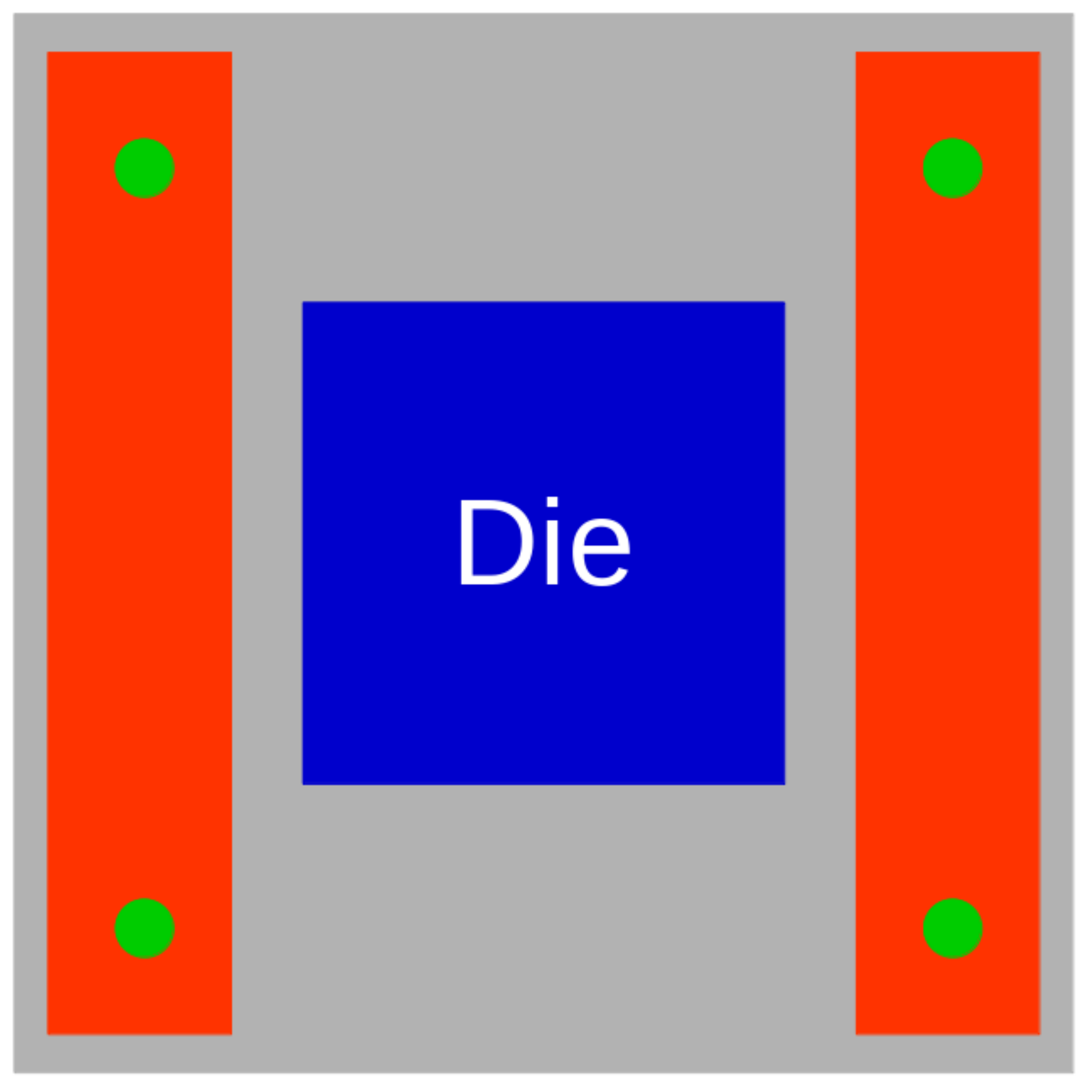}
		\caption{}
		\label{fig:LED_structure_TV}
	\end{subfigure}
	\qquad
	~ 
	\begin{subfigure}[b]{0.15\textwidth}
		\includegraphics[width=\textwidth]{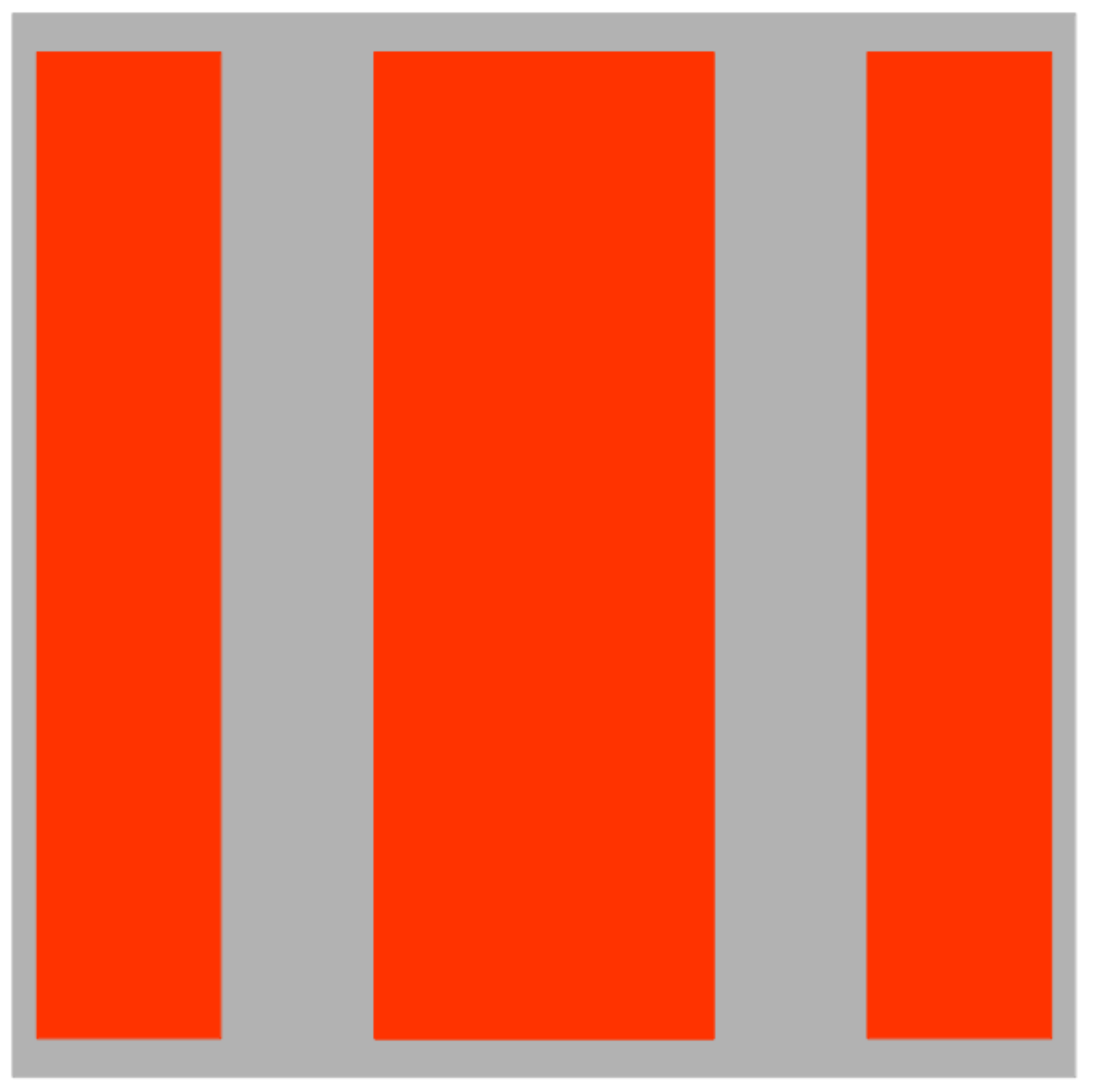}
		\caption{}
		\label{fig:LED_structure_BV}
	\end{subfigure}
		\caption{(\subref{fig:LED_structure_CSV}) LED cross-section view. (\subref{fig:LED_structure_TV}) LED top view without epoxy dome. (\subref{fig:LED_structure_BV}) LED bottom view. Drawings not to scale.}\label{fig:structure_led_cree}
\end{figure}

The thermal analysis was performed for the package of Cree\textregistered\space white XLamp\textregistered\space XB-D LEDs. These devices have a small footprint (2.45$\times$2.45~mm\textsuperscript{2}), a maximum current rating of 1~A and are available in different colors (white, blue, green, amber, and red). The cross-section schematic view of the LEDs can be seen in Fig.~\ref{fig:LED_structure_CSV}~\cite{Cree2015,Note2011}. According to information provided by the manufacturer, the LED die (in blue) is attached to an AlN carrier (in grey) and the junction terminals are electrically connected to the Cu electrodes (red shapes) via bond wires. The die is composed of a silicon carbide (SiC) substrate~\cite{Digi-Key2015} with the GaN active layers~\cite{Digi-Key2013} on top. Fig.~\ref{fig:LED_structure_TV} shows the top view of the LED without the epoxy dome, with the electrode pads (red shapes) and vias holes (green circular shapes) clearly visible. Fig.~\ref{fig:LED_structure_BV} shows the bottom view of the LED; the electrode in the middle corresponds to the thermal pad. The dimensions and properties of each part are listed in Table~\ref{table:LED_parts_and_properties}.

To evaluate the impact of the carrier on the reliability of the LED, the simulations were performed for different materials. In addition to Al\textsubscript{2}O\textsubscript{3}, AlN, and Si (materials typically used for this purpose), diamond carriers were also considered. Al\textsubscript{2}O\textsubscript{3} and AlN are available in ceramic and crystalline forms, with significantly different values of $\kappa$. In order to account for this variation, the simulations were performed with the minimum and maximum values of $\kappa$ found in the literature for the ceramic and crystalline forms~\cite{Liu2011,Karditsas,Kyocera2015,Materialsa,Huang2019,Slack1987,Inc.}. Diamond is also available as a single crystal (SCD) or as a polycrystalline (PCD) material~\cite{Materials,DiamondMaterials2014}; both types were considered in the simulations. The $\kappa$ of the different materials is listed in Table~\ref{table:LED_parts_and_properties}.\\

\begin{table*}[htb]
\begin{center}
	\begin{tabular}{ccccc}\toprule
		\bf\multirow{2}{*}{Part} 	& \bf\multirow{2}{*}{Material} 			& \bf Thickness 	& \bf Area 					& \bf Thermal Conductivity  \\
		~							& ~ 		  							& $\mu$m 	        & mm\textsuperscript{2} 	& W/(m$\cdot$K)\\  
		\midrule
		\bf{Carrier}	 			& Al\textsubscript{2}O\textsubscript{3} $^{1)}$ 
																			& 540     			& 2.45$\times$2.45			& 27~\cite{Liu2011} / 35~\cite{Karditsas}    \\
	
		~	 			            & Al\textsubscript{2}O\textsubscript{3} $^{2)}$ 
											 								&  ~     			& ~           				& 42~\cite{Kyocera2015} \\
		~			 				& AlN $^{1)}$ 
		                                                                    & ~		 			& ~							& 60~\cite{Materialsa} / 193~\cite{Huang2019} \\
		~				 			& AlN $^{2)}$ 
		                                                                    &  ~     			& ~      					& 285~\cite{Slack1987} \\											 								
		~				 			& Si $^{2)}$ 
		                                                                    &  ~     			& ~           				& 148~\cite{Inc.}	\\		
		~				 			& Diamond $^{3)}$ 
		                                                                    &  ~     			& ~           			    & 1800~\cite{Materials}\\	
		~				 			& Diamond $^{2)}$ 
		                                                                    &  ~     			& ~           				& 2200~\cite{DiamondMaterials2014} \\
		\hdashline[0.5pt/5pt]
		\bf{Led die}				& GaN			     					& 5                 & 1.06$\times$1.06  		& 230~\cite{Mion2006} \\
		~						 	& SiC 	  								& 140      		    & 1.06$\times$1.06   		& 430~\cite{Wolfspeed2020} \\
		\hdashline[0.5pt/5pt]
		\bf{Top electrodes}			& Cu 			     					& 110         		& 2.29$\times$0.36 			& 400~\cite{Edge}  \\
		\hdashline[0.5pt/5pt]
		\bf{Bottom electrodes}		& Cu      								& 110       		& 2.29$\times$0.33 			& 400~\cite{Edge}  \\
		\hdashline[0.5pt/5pt]
		\bf{Bottom thermal pad}		& Cu      								& 110        		& 2.29$\times$0.92 			& 400~\cite{Edge}  \\
		\hdashline[0.5pt/5pt]
		\bf{Vias holes}				& Solder  			    				& 540          		& 4.07$\times$10\textsuperscript{-3}
																															& 48~\cite{Edge}  \\
		\hdashline[0.5pt/5pt]
		\bf{Epoxy}					& Silicone								& 220 $^{4)}$ / 1090 $^{6)}$  	
																								& 4~mm\textsuperscript{3} $^{5)}$ 
																												 			& $0.3$~\cite{Tsai2012} \\

		\bottomrule
	\end{tabular}
	\end{center}
	\footnotesize{$^{1)}$ ceramic; $^{2)}$ single crystal; $^{3)}$ polycrystalline; $^{4)}$ rectangle thickness; $^{5)}$ volume; $^{6)}$ dome radius.}
	\caption{Dimensions and thermal conductivity of LED parts}
	\label{table:LED_parts_and_properties}
\end{table*}

\subsection{Numerical simulations}\label{simulations_description}

The simulations were performed using Ansys multiphysics software (package Mechanical). The structure of the LED was created using the information from manufacturer as in Table~\ref{table:LED_parts_and_properties}; the die-attach layer was neglected and the thermal contact between the GaN+SiC die and the carrier was considered to be perfect. 

After being generated, the mesh was validated using Ansys built-in validation tools. An element quality measure lower than 5$\times$10\textsuperscript{-6} was used; this metric is based on the ratio of the volume to the sum of the squares of the edge lengths for the three-dimensional (3D) mesh elements, which, according to the Ansys Meshing User's guide, has proven to be effective for thermal problems. 

The thermal power $P_{\rm Th}$, calculated as 75\% of the electric power $P_{\rm El}$~\cite{Note2015a}, was considered to be generated inside the 5-$\mu$m-thick GaN layer and the temperature of the three bottom electrodes was kept at 40\textdegree C. This assumption corresponds to mounting the LED directly on a TEC; even though this may not correspond to a realistic scenario, it "forces" the transfer of heat across the different components of the LED and allows a more direct evaluation of the impact of the $\kappa$ of the carrier on $T_{\rm J}$. The heat generated within the active layers of the die was also considered to be dissipated by convection from the epoxy and from the lateral sides of the carrier; the convective film coefficient was the default value assumed by Ansys (5~W/(m$^2\cdot$K)).

The simulations were performed for current levels $I$ between 100 and 800~mA with a step of 100~mA and for 350~mA (nominal current value according to the manufacturer). The electric power $P_{\rm El}$ was calculated as the product of the diode forward current $I$ and the forward voltage $V$, which, in turn, were determined from the LTSpice model for the Xamp\textregistered\space XB-D LED~\cite{Cree}. \\

\section{Results and discussion}

\subsection{\label{Junction_temperature}Impact of the carrier on the junction temperature}

\begin{figure}[t]
	\centering
	\begin{subfigure}[b]{0.35\textwidth}
		\includegraphics[width=\textwidth]{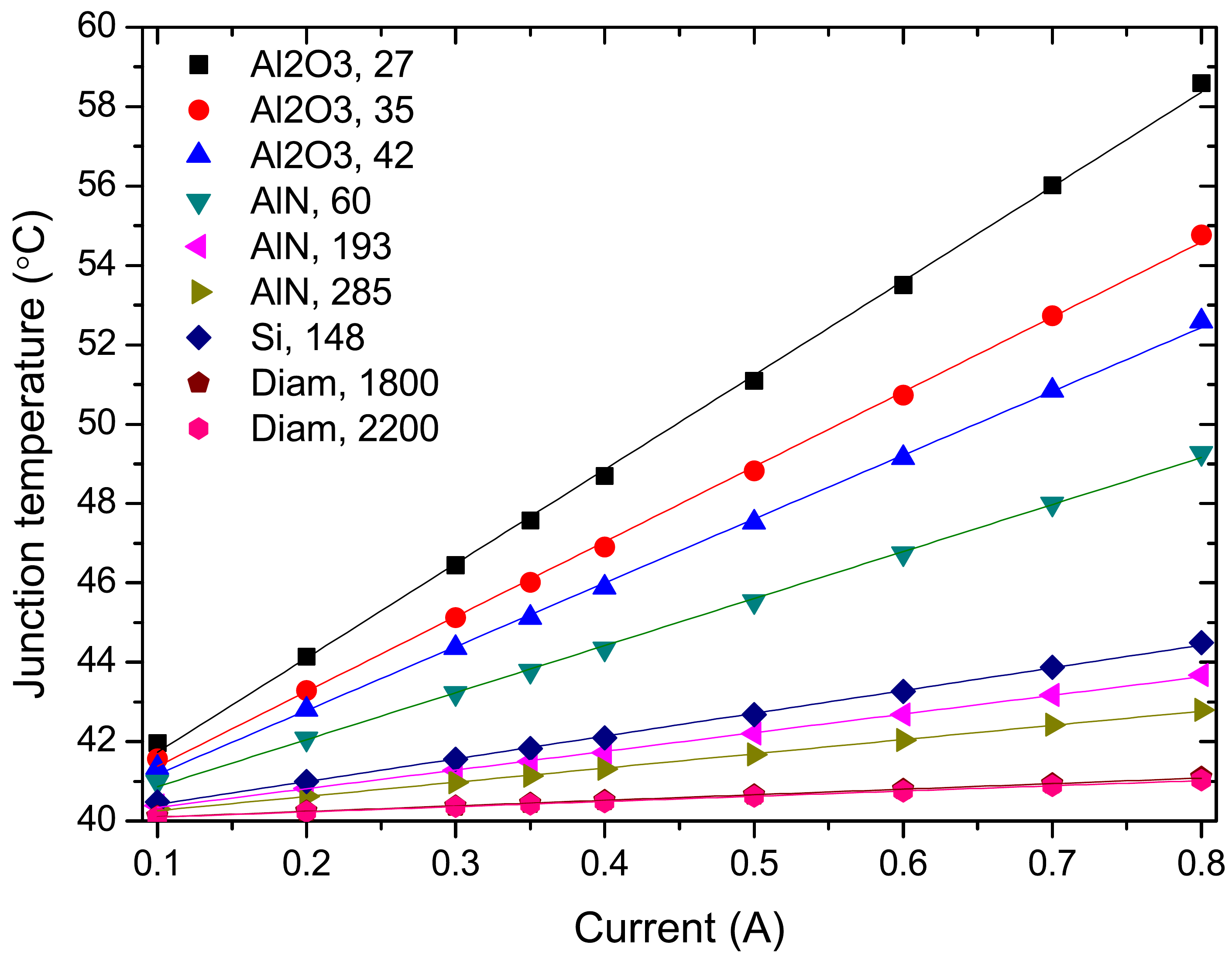}
		\caption{}
		\label{fig:Junct_temp}
	\end{subfigure}
	\qquad
	\begin{subfigure}[b]{0.35\textwidth}
		\includegraphics[width=\textwidth]{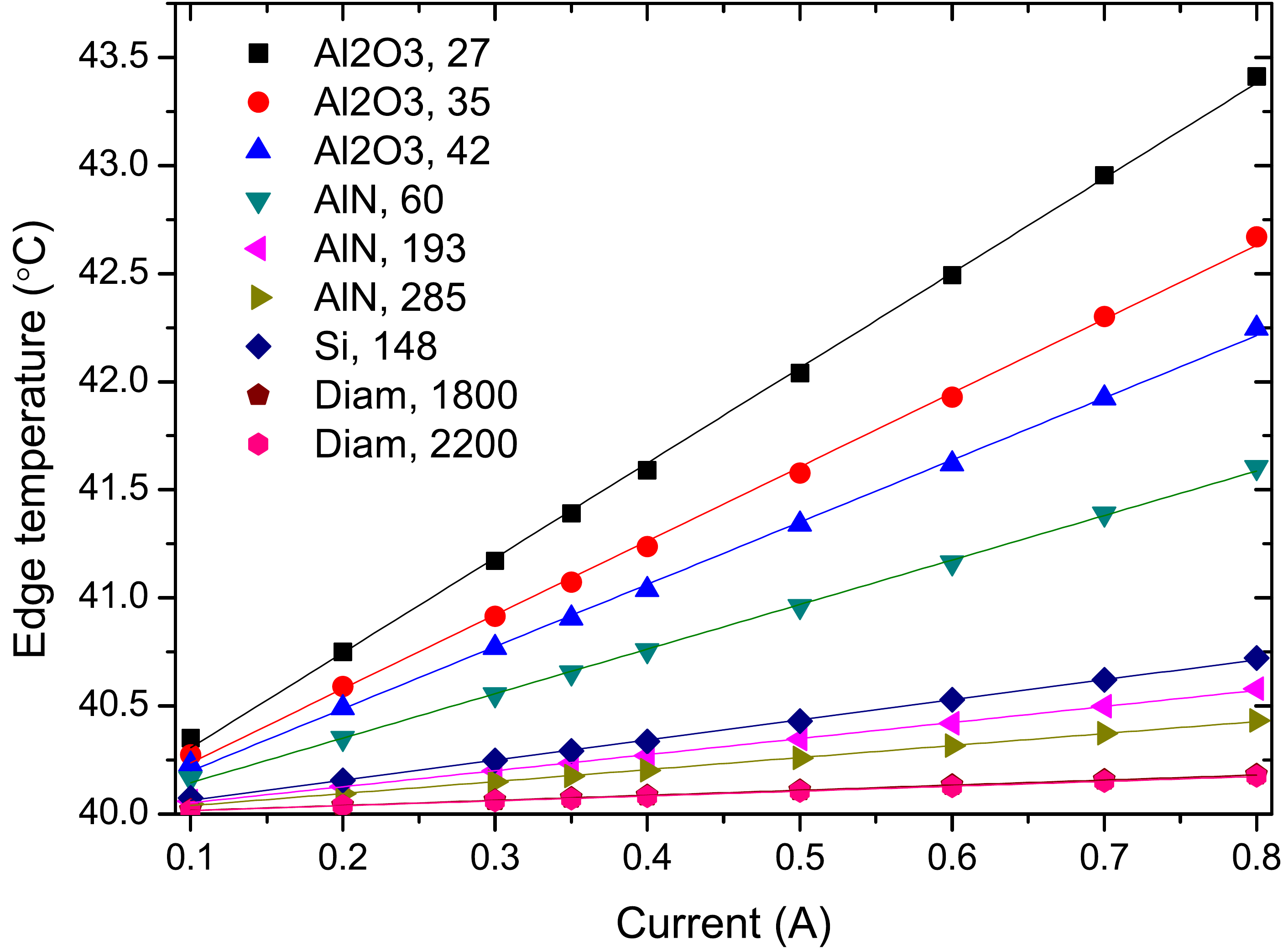}
		\caption{}
		\label{fig:Edge_temp}
	\end{subfigure}
	\qquad
	\begin{subfigure}[b]{0.35\textwidth}
		\includegraphics[width=\textwidth]{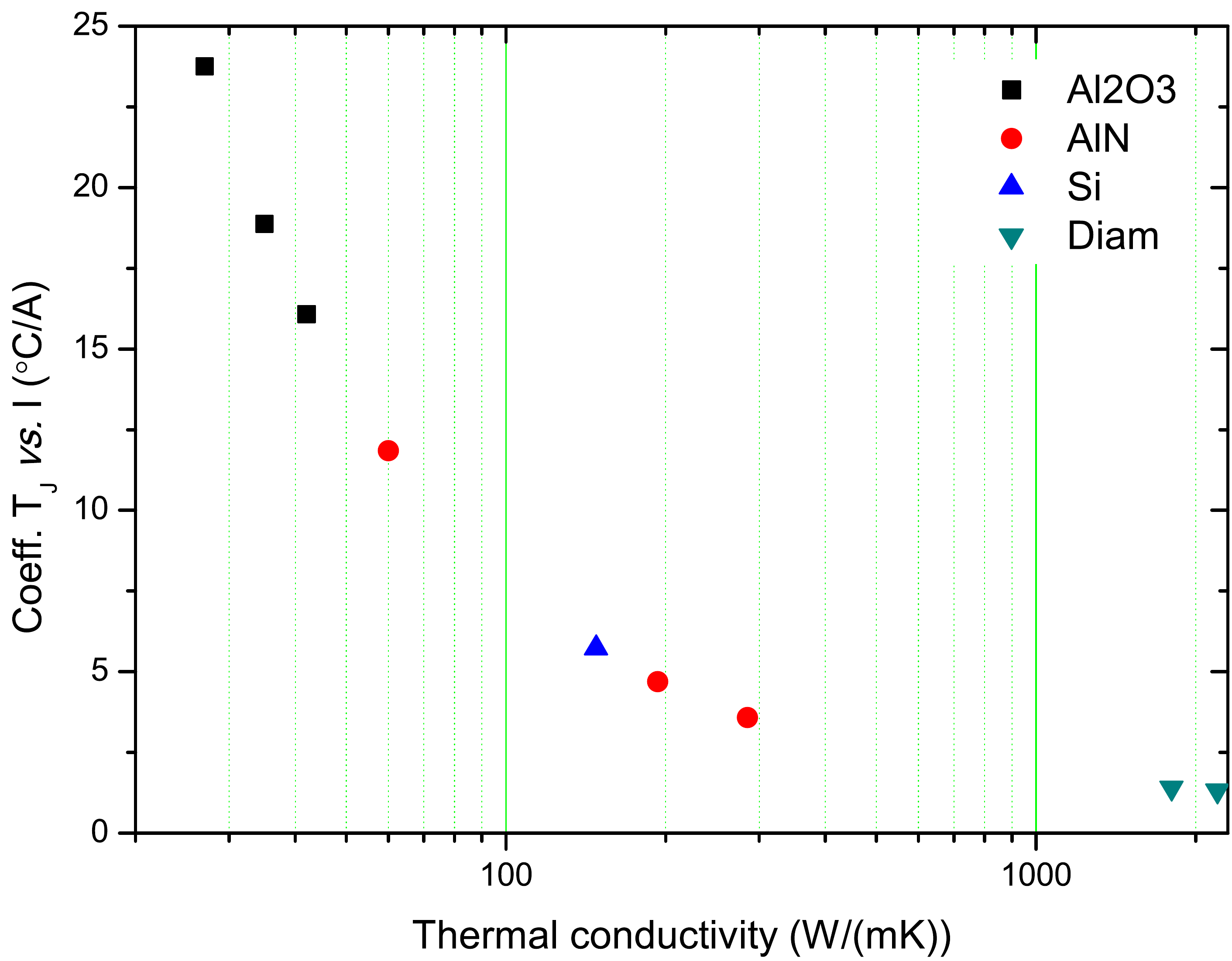}
		\caption{}
		\label{fig:Coeff_junct_temp_current}
	\end{subfigure}	
	\caption{(\subref{fig:Junct_temp}) $T_{\rm J}$ and respective trend lines as a function of $I$. (\subref{fig:Edge_temp}) Edge temperature as a function of $I$. (\subref{fig:Coeff_junct_temp_current}) Slope of $T_{\rm J}(I)$ curves as a function of the $\kappa$ of the carrier.}\label{fig:Junct_edge_temp}
\end{figure}

The simulated values of $T_{\rm J}$ are plotted as a function of $I$ for each carrier in Fig.~\ref{fig:Junct_temp} with solid symbols. The respective trend lines are also represented as solid lines. 
As expected, $T_{\rm J}$ decreases with the $\kappa$ of the carrier. It is clear that not only the type of material (Al\textsubscript{2}O\textsubscript{3}, AlN, Si or diamond) but also the quality (single crystal vs ceramic) have a large impact on the junction temperature. As an example, if the lowest quality Al\textsubscript{2}O\textsubscript{3} is used as the carrier, $T_{\rm J}$ increases by 6\textdegree C with respect to $T_{\rm J}$ obtained with the single crystal form. A similar situation happens with AlN; $T_{\rm J}$ increases by $\simeq$6.6\textdegree C with the lowest quality material. Regarding Si, the simulations were performed considering only the crystalline material with a $\kappa$=148~W/(m$\cdot$K). Ceramic AlN can have a higher (193~W/(m$\cdot$K)) or lower (60~W/(m$\cdot$K)) $\kappa$, as a consequence the $T_{\rm J}$ obtained with Si may be higher or lower than then the one obtained with AlN, depending on the quality of the latter. Replacing crystalline AlN with PCD ($\kappa$=1800~W/(m$\cdot$K)) reduces $T_{\rm J}$ by $\simeq$1.7\textdegree C; the use of SCD only brings a further $\simeq$0.1\textdegree C benefit. 

\begin{figure}[t]
	\centering
	\begin{subfigure}[b]{0.35\textwidth}
		\includegraphics[width=\textwidth]{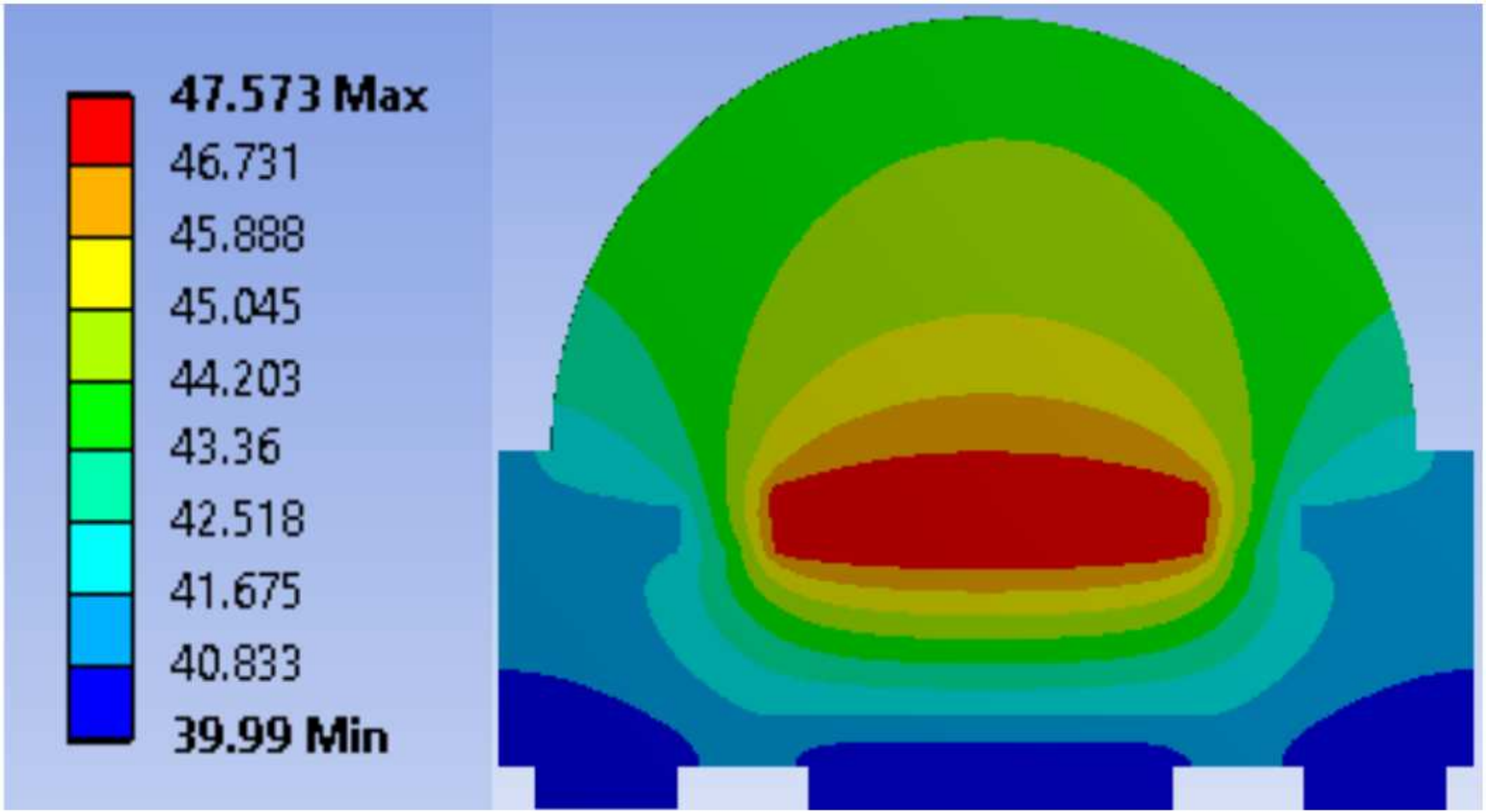}
		\caption{}
		\label{fig:350_TempMap_Alumina_27_CSV}
	\end{subfigure}	
	\qquad
	\begin{subfigure}[b]{0.35\textwidth}
		\includegraphics[width=\textwidth]{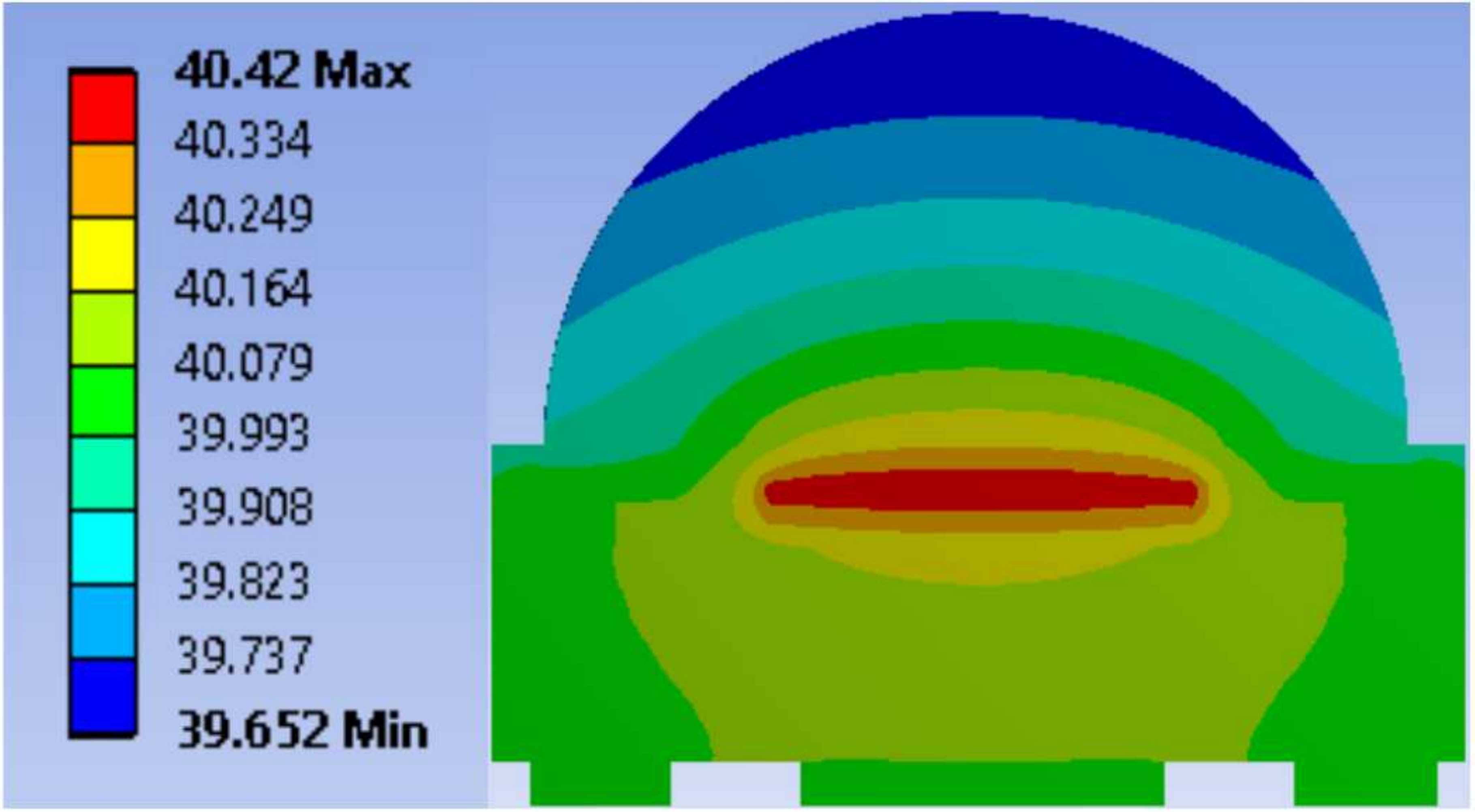}
		\caption{}
		\label{fig:350_TempMap_Diam_2200_CSV}
	\end{subfigure}	
	\caption{Cross-section temperature maps at 350~mA with (\subref{fig:350_TempMap_Alumina_27_CSV}) Al\textsubscript{2}O\textsubscript{3} (27~W/(m$\cdot$K)) and (\subref{fig:350_TempMap_Diam_2200_CSV}) diamond (2200~W/(m$\cdot$K)).
	}\label{fig:Temperature_maps_350mA}
\end{figure}

\begin{figure}
	\centering
	\begin{subfigure}[b]{0.35\textwidth}
		\includegraphics[width=\textwidth]{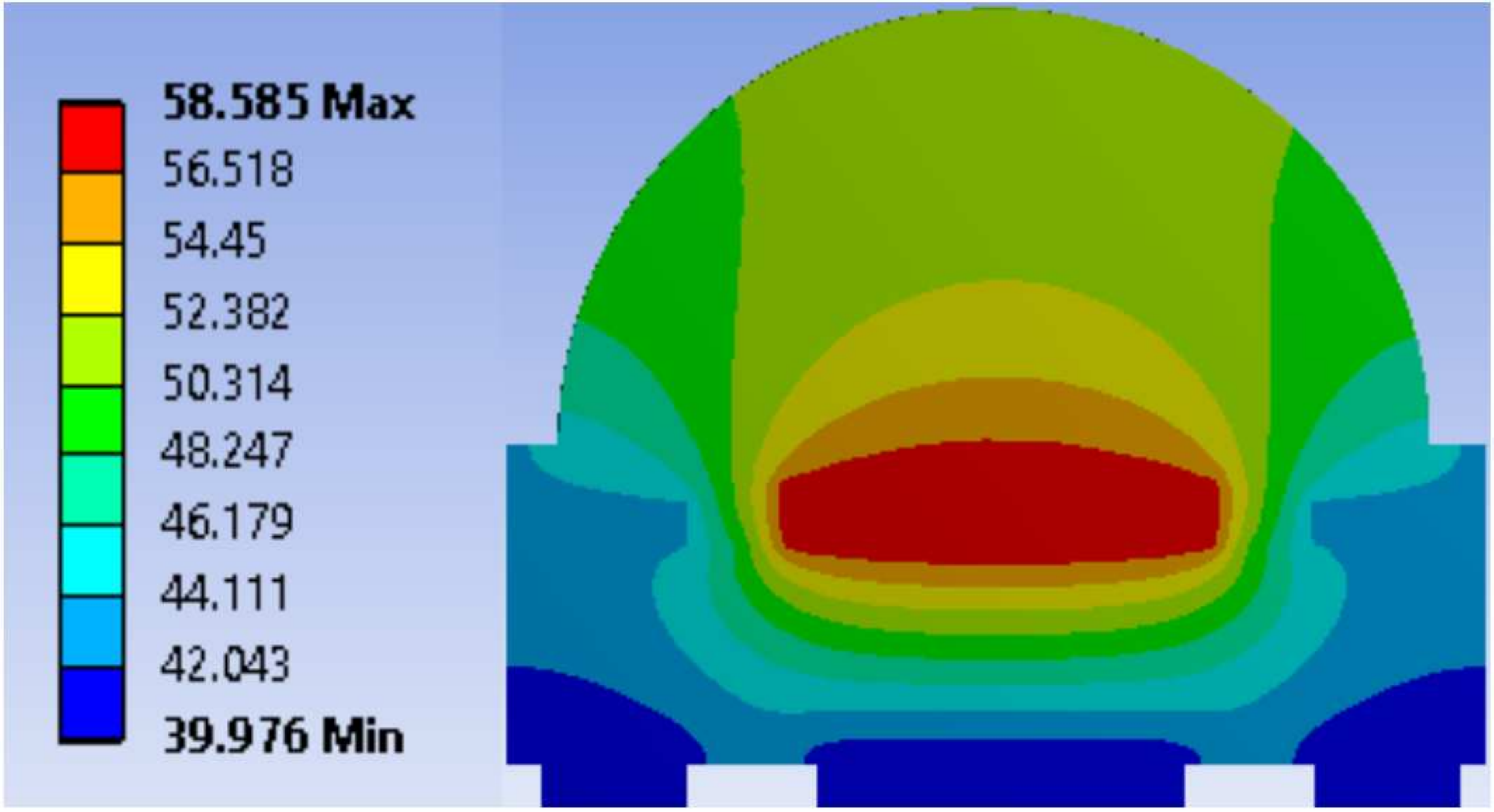}
		\caption{}
		\label{fig:800_TempMap_Alumina_27_CSV}
	\end{subfigure}	
    \qquad
    \begin{subfigure}[b]{0.35\textwidth}
    	\includegraphics[width=\textwidth]{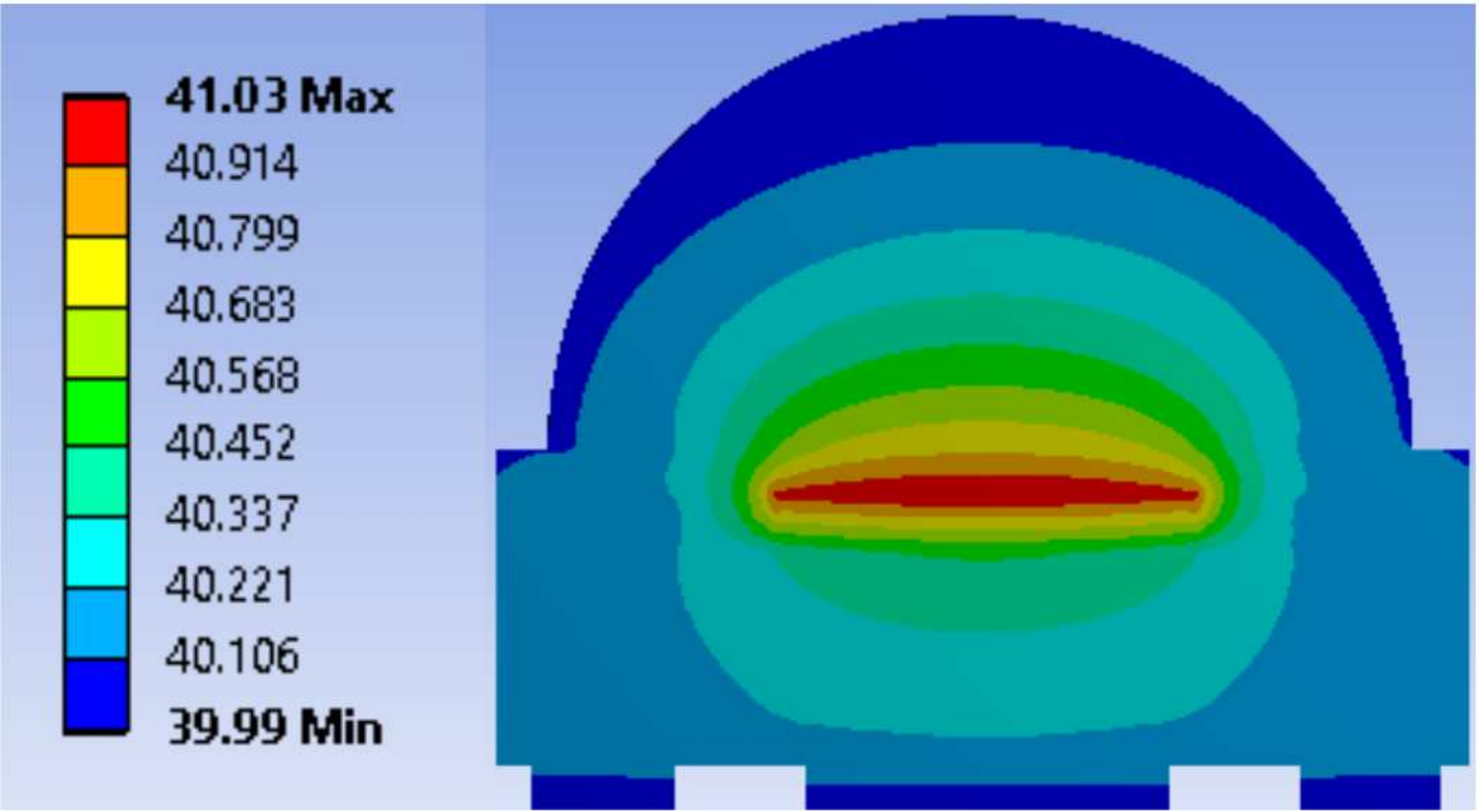}
    	\caption{}
    	\label{fig:800_TempMap_Diam_2200_CSV}
    \end{subfigure}	
	\caption{Cross-section temperature maps at 800~mA with  (\subref{fig:800_TempMap_Alumina_27_CSV}) Al\textsubscript{2}O\textsubscript{3} (27~W/(m$\cdot$K) and (\subref{fig:800_TempMap_Diam_2200_CSV}) diamond (2200~W/(m$\cdot$K)).
	}\label{fig:Temperature_maps_800mA}
\end{figure}

The temperature maps obtained with the two extreme materials (Al\textsubscript{2}O\textsubscript{3}, 27~W/(m$\cdot$K) and diamond, 2200~W/(m$\cdot$K)) at 350 or 800~mA are shown in Figs.~\ref{fig:Temperature_maps_350mA} and \ref{fig:Temperature_maps_800mA}, respectively. The difference in $T_{\rm J}$ is notorious: for nominal current (350~mA), $T_{\rm J}$ reaches 47.6 and 40.4\textdegree C with Al\textsubscript{2}O\textsubscript{3} and SCD, respectively ($\simeq$7.2\textdegree C difference). Similarly, if the current raises to 800~mA, $T_{\rm J}$ increases to 58.6 and 41.0\textdegree C ($\simeq$17.6\textdegree C difference) for the same materials. Independently of the current level, the SCD carrier behaves nearly as a thermal short circuit, keeping $T_{\rm J}$ close to the temperature imposed by the TEC.

The impact of the carrier on $T_{\rm J}$ can also be evaluated by calculating the slope of the $T_{\rm J}(I)$ curves, which is depicted in Fig.~\ref{fig:Coeff_junct_temp_current} as a function of the $\kappa$ of the carrier. A lower slope value indicates a smaller dependency of the $T_{\rm J}$ on the current level. The slope varies considerably with the quality of Al$_2$O$_3$ (between $\simeq$16 and $\simeq$24\textdegree C/A) and AlN (between $\simeq$3.6 and $\simeq$12\textdegree C/A) and reaches its minimum value with the diamond carriers. No significant difference is observed between SCD and PCD materials ($\simeq$1.4\textdegree C/A).

It should be mentioned that the apparently linear dependence of $T_{\rm J}$ with $I$ is a consequence of the assumptions made in the implementation of the simulations. The first assumption is the validity of the LTSpice model used to calculate the voltage drop across the LED for the different levels of $I$. This model is valid for $T_{\rm J}$=25\textdegree C, whereas in our case the values of $T_{\rm J}$ are higher. In addition, for a given $I$, $T_{\rm J}$ varies with the carrier. Since the LED voltage also depends on $T_{\rm J}$, the use of this simple LTSpice model introduces an error in the determination of the electric power $P_{\rm El}$ for a given $I$. Nevertheless, for this particular model, an increase of 10 times (900\%) in the current is accompanied by a change in the LED voltage smaller than 20\%, so this is not considered to be the main source of error. The most relevant assumption is the consideration that $P_{\rm Th}$ is 75\% of $P_{\rm El}$ for all levels of $I$. The 25\% efficiency is valid for nominal operating conditions~\cite{Note2015a}. As the current increases, the efficiency of the LED decreases, so the dissipated power (and correspondingly $T_{\rm J}$) increases. However, and despite this assumption contributes with an error that may not be negligible, taking into account the temperature-dependent efficiency would increase the difference in the thermal performance of the carriers even more. By assuming a constant efficiency, we are indirectly imposing the lower limit of the difference in the performance of the different carriers. The impact of considering a temperature-dependent efficiency will be a part of our future work.

The carrier also influences the thermal footprint of the LED die. This can be seen in Fig. \ref{fig:Edge_temp}, that shows the temperature at the edge of the carrier for the different current levels. With  Al\textsubscript{2}O\textsubscript{3} (27~W/(m$\cdot$K)), this temperature is as high as 41.4 and 43.4\textdegree C for 350 and 800~mA, respectively, whereas the temperature with diamond (2200~W/(m$\cdot$K)) at the same location stays at $\simeq$40.1 and $\simeq$40.2\textdegree C for the same current levels, respectively. The values of temperature at the edge of the carrier obtained with the other materials are between these extreme values. The impact of the carrier on the device footprint becomes even more apparent if one looks at the cross-section temperature maps obtained with a current level of 350 and 800~mA for the two extreme materials (Figs. \ref{fig:Temperature_maps_350mA} and \ref{fig:Temperature_maps_800mA}). Figs. \ref{fig:350_TempMap_Alumina_27_CSV} and \ref{fig:800_TempMap_Alumina_27_CSV} show the results obtained with Al\textsubscript{2}O\textsubscript{3}, 27~W/(m$\cdot$K)  and  Figs. \ref{fig:350_TempMap_Diam_2200_CSV} and \ref{fig:800_TempMap_Diam_2200_CSV} show the results obtained with diamond, 2200~W/(m$\cdot$K). With the diamond carrier, the temperature at the immediate vicinity of the die is kept extremely close to the value imposed by the TEC.

\subsection{\label{LED_characteristics}Impact of the carrier on the LED characteristics}

$T_{\rm J}$ impacts directly the LED characteristics such as emission intensity, lifetime, and stability of the wavelength. The variation of each parameter with $T_{\rm J}$ is analysed in detail in the following paragraphs. To facilitate the interpretation of the results, the impact of the carrier on the emission intensity and lifetime is evaluated considering the emission intensity and lifetime obtained with AlN and $\kappa$=193~W/(m$\cdot$K) as a reference.

Similarly to other p-n junctions, the current that flows through an LED increases with increasing temperature due to (among other factors) the dependency of the carrier concentration on the temperature given by the Boltzmann equations
\begin{equation}
	n=N_{\rm C}\cdot\exp\left({-\frac{E_{\rm C}-E_{\rm F}}{k_{\rm B}\cdot T}}\right)
	\label{eq:Boltzmann_electrons}
\end{equation} 
and
\begin{equation}
    p=N_{\rm V}\cdot\exp\left(\frac{E_{V}-E_{\rm F}}{k_{\rm B}\cdot T}\right)
	\label{eq:Boltzmann_holes}
\end{equation} 
 for electrons and holes, respectively. In these equations, $N_{\rm C}$ and $N_{\rm V}$ are the impurity densities in the n and p sides of the junction, respectively, $E_{\rm C}$ and $E_{\rm V}$ are the bottom and top of the conduction and valence bands, respectively, $E_{\rm F}$ is the Fermi level, $k_{\rm B}$ is the Boltzmann constant and $T$ is the junction temperature. However, in the case of LEDs, the temperature-induced increase of the current does not translate to an increased emission intensity because, as temperature rises, non-radiant recombination (such as Shockley Read Hall and Auger recombination) increases, whereas radiative recombination (the source of the LED light emission) decreases. Carrier loss over the heterostructure barriers, which also contributes to the decrease of the LED emission efficiency, also increases with temperature. For more details on the related physical phenomena the readers are referred to works such as~\cite{Schubert2006}. 

\begin{figure}
	\centering
	\begin{subfigure}[b]{0.45\textwidth}
		\includegraphics[width=\textwidth]{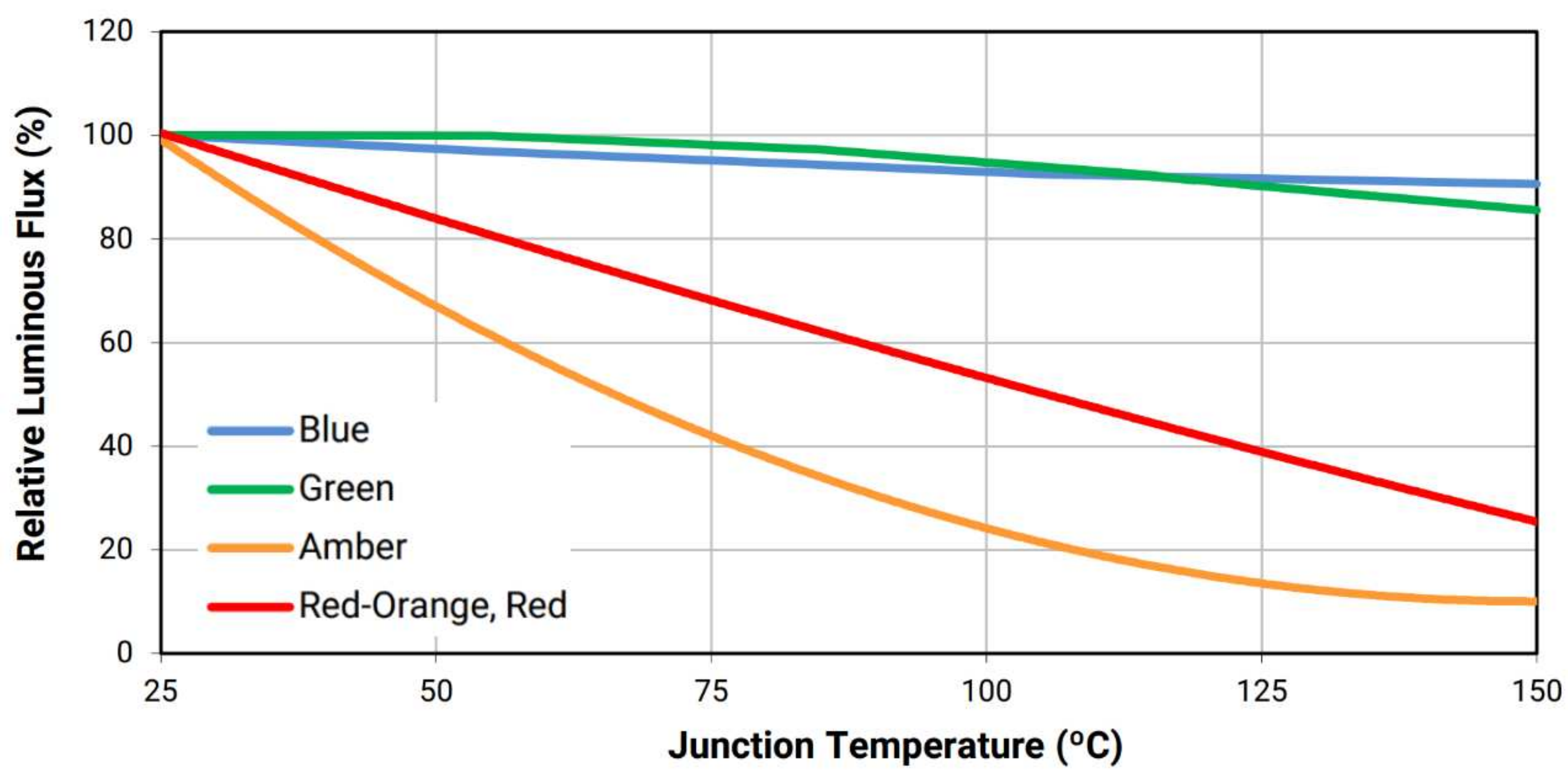}
		\caption{}
		\label{fig:RLF_temp}
	\end{subfigure}
	\qquad
	\begin{subfigure}[b]{0.4\textwidth}
		\includegraphics[width=\textwidth]{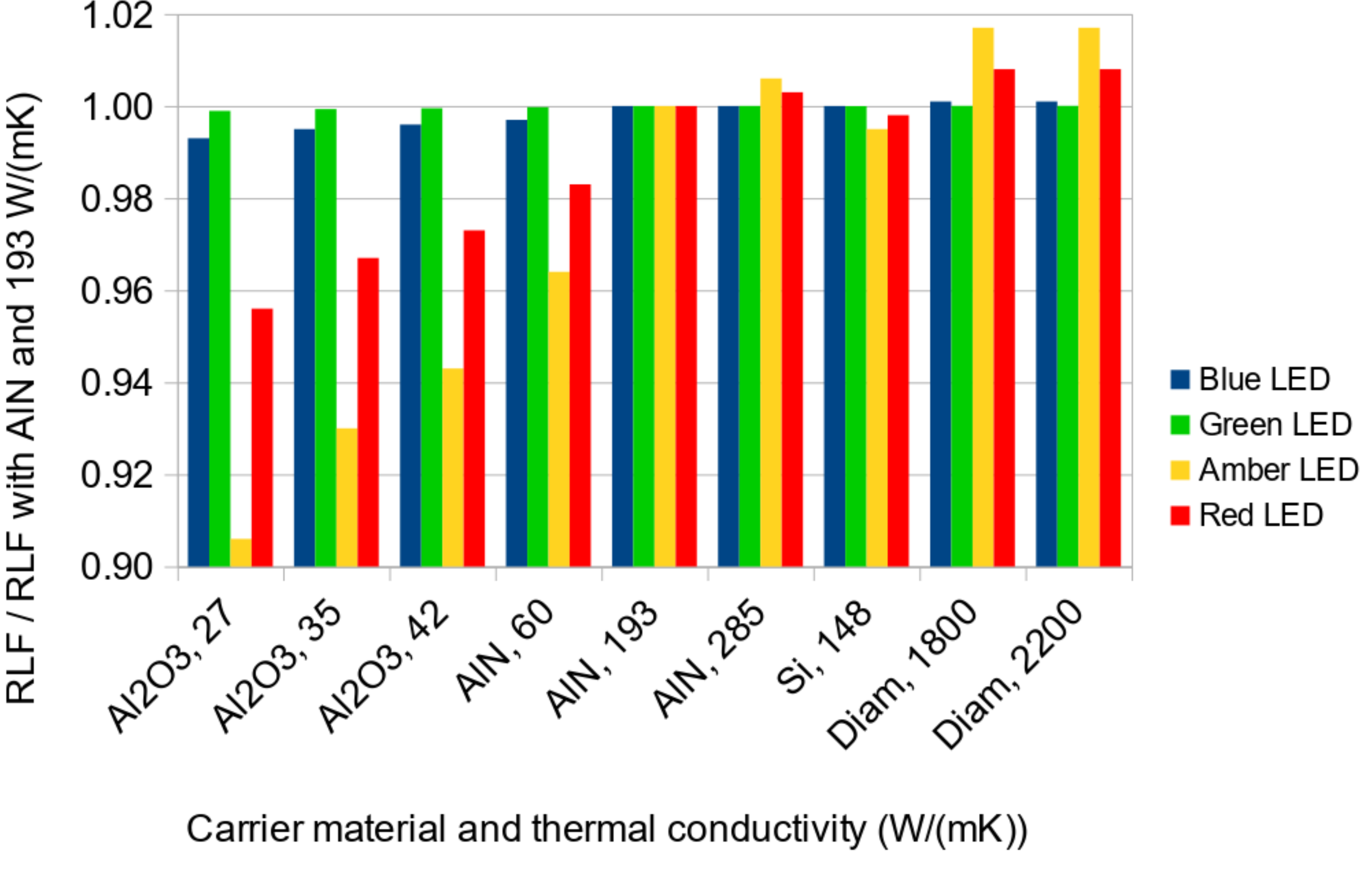}
		\caption{}
		\label{fig:RLF}
	\end{subfigure}
	\qquad
	\caption{(\subref{fig:RLF_temp}) RLF as a function of $T_{\rm J}$ for 350~mA. Image obtained from the device datasheet~\cite{Cree2015}. (\subref{fig:RLF}) Relative variation of RLF with respect to the values obtained with AlN ($\kappa$=193~W/(m$\cdot$K).}\label{fig:RLF_Figures}
\end{figure}

Near room temperature the  dependence of the emission intensity on the temperature may be described by the phenomenological equation: 
\begin{equation}
	I=I_{300\rm K}\cdot\exp\left(-\frac{T-300}{T_{1}}\right),
	\label{eq:Emission_intensity}
\end{equation} 
where $I_{300\rm K}$ is the emission intensity at 300~K and $T_{1}$ is the characteristic temperature, an empirical parameter that depends on the  heterojunction layers of a particular device~\cite{Schubert2006}. In the current case, however, this equation was of no use since the manufacturer does not give details about the structure of the active layers of the LED - which prevents the determination of the characteristic temperature. Instead, the impact of the carrier material on the emission intensity was estimated using the characteristic curves representing the RLF as a function of $T_{\rm J}$ provided by the manufacturer for a current of 350~mA~\cite{Cree2015} (Fig.~\ref{fig:RLF_temp}). The dependency of $T_{\rm J}$ on $I$ was considered independent of the LED colour; under this assumption, the values of $T_{\rm J}$ obtained with each carrier at 350~mA can be used to estimate the value of the RLF for each type of LED using the respective curve in Fig.~\ref{fig:RLF_temp}. The ratio between the RLF obtained at 350~mA with each carrier and the RLF obtained with an AlN carrier ($\kappa$=193~W/(m$\cdot$K)) is represented in Fig.~\ref{fig:RLF} for the blue, green, amber, and red LEDs. The green LED does not show a measurable variation of the RLF with the carrier, whereas the RLF of the amber LED increases (or decreases) by 2\% (by 10\%) when AlN is replaced with diamond 2200~W/(m$\cdot$K) (Al\textsubscript{2}O\textsubscript{3}, 27~W/(m$\cdot$K)).

The carrier also influences the lifetime directly. It is generally accepted by the semiconductor industry to determine the Time-to-Failure (TTF) of a device using the Black model~\cite{Black1969,Bahl2015}:
\begin{equation}
	TTF=I_{0}\cdot J^{-n}\cdot\exp\left(\frac{q\cdot E_{\rm a}}{k_{\rm B}\cdot T}\right),
	\label{eq:TTF}
\end{equation} 
where $I_{0}$ is a constant, $J$ is the current density, $n$ is a scaling factor, $q$ is the electron charge, $E_{\rm a}$ is the activation energy of the failure mechanisms (in eV), $k_{\rm B}$ is the Boltzmann constant, and $T$ is the junction temperature (in K). As  expected, the TTF decreases as the junction temperature increases. Typically manufacturers do not provide the numerical values of the constants $I_0$, $n$, and $E_a$, so the lifetime of the device cannot be calculated directly. Instead, it is common to consider the lifetime under nominal operating conditions, and to evaluate the impact of elevated operating temperatures on the lifetime by calculating the so-called acceleration factor (AF). The AF based on the Black model can be defined as:
\begin{equation}
	AF=
	\frac{TTF_{\rm nom}}{TTF_{\rm st}}=
	\left(
	  \frac{J_{\rm st}}{J_{\rm nom}}
	\right)^n\cdot
	e^{
	\frac{q\cdot E_{\rm a}}{k_{\rm B}}\cdot
	\left(
	  \frac{1}{T_{\rm nom}}-\frac{1}{T_{\rm st}}
	\right)},
	\label{eq:AF_Black}
\end{equation}
where $J_{\rm nom}$ (and $T_{\rm nom}$) and $J_{\rm st}$ (and  $T_{\rm st}$) are the current density levels (junction temperature) at the nominal and stress conditions, respectively, and $E_{\rm A}$ is  the activation energy of the failure mechanisms of the semiconductor material. Defined this way, the AF correlates the actual high temperature operating life (HTOL) stress test data points, taken at elevated temperatures and/or current levels, to the expected lifetime under the actual operating conditions in a given application.

\begin{figure}
	\centering
	\begin{subfigure}[b]{0.35\textwidth}
		\includegraphics[width=\textwidth]{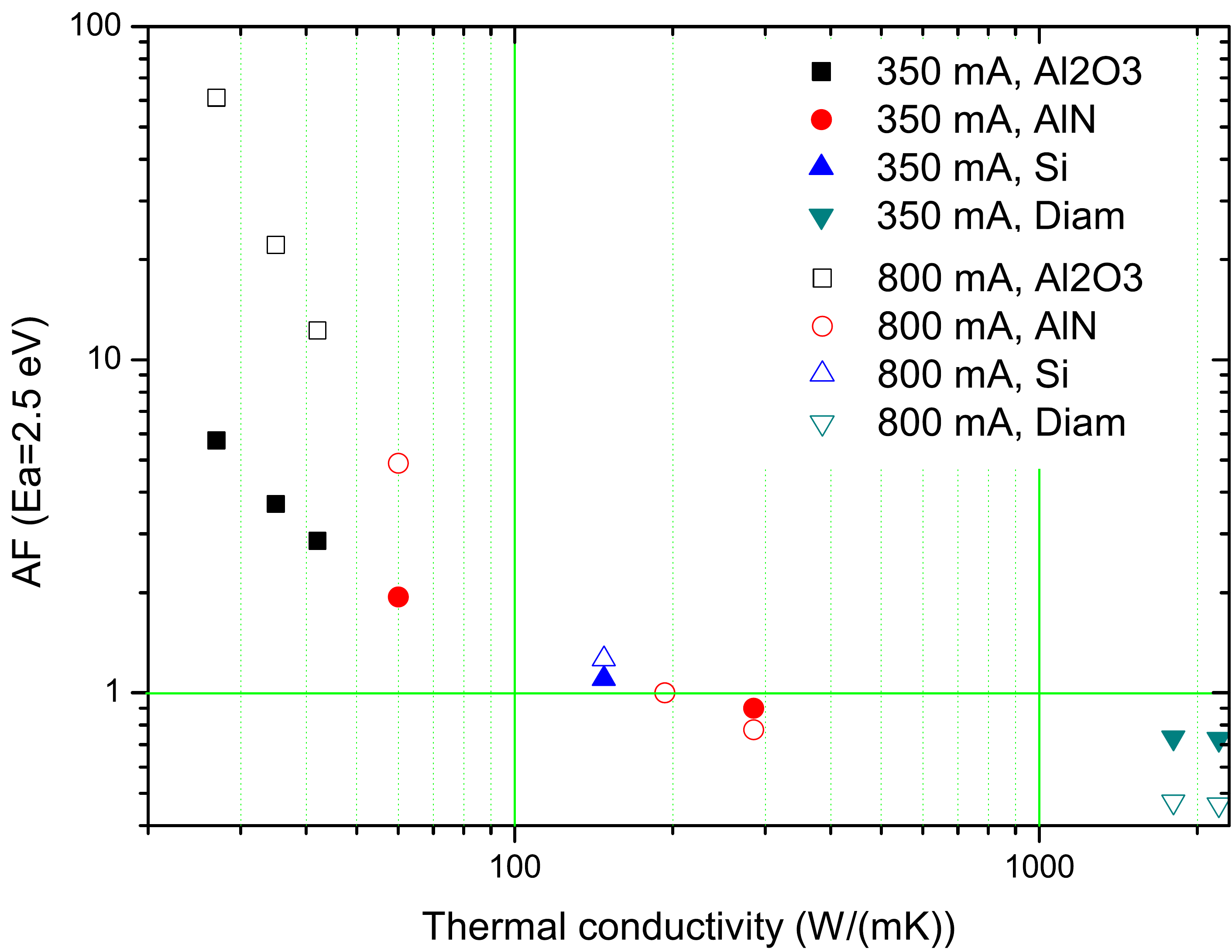}
		\caption{}
		\label{fig:AF_2.5eV}
	\end{subfigure}
	\qquad
	\begin{subfigure}[b]{0.35\textwidth}
		\includegraphics[width=\textwidth]{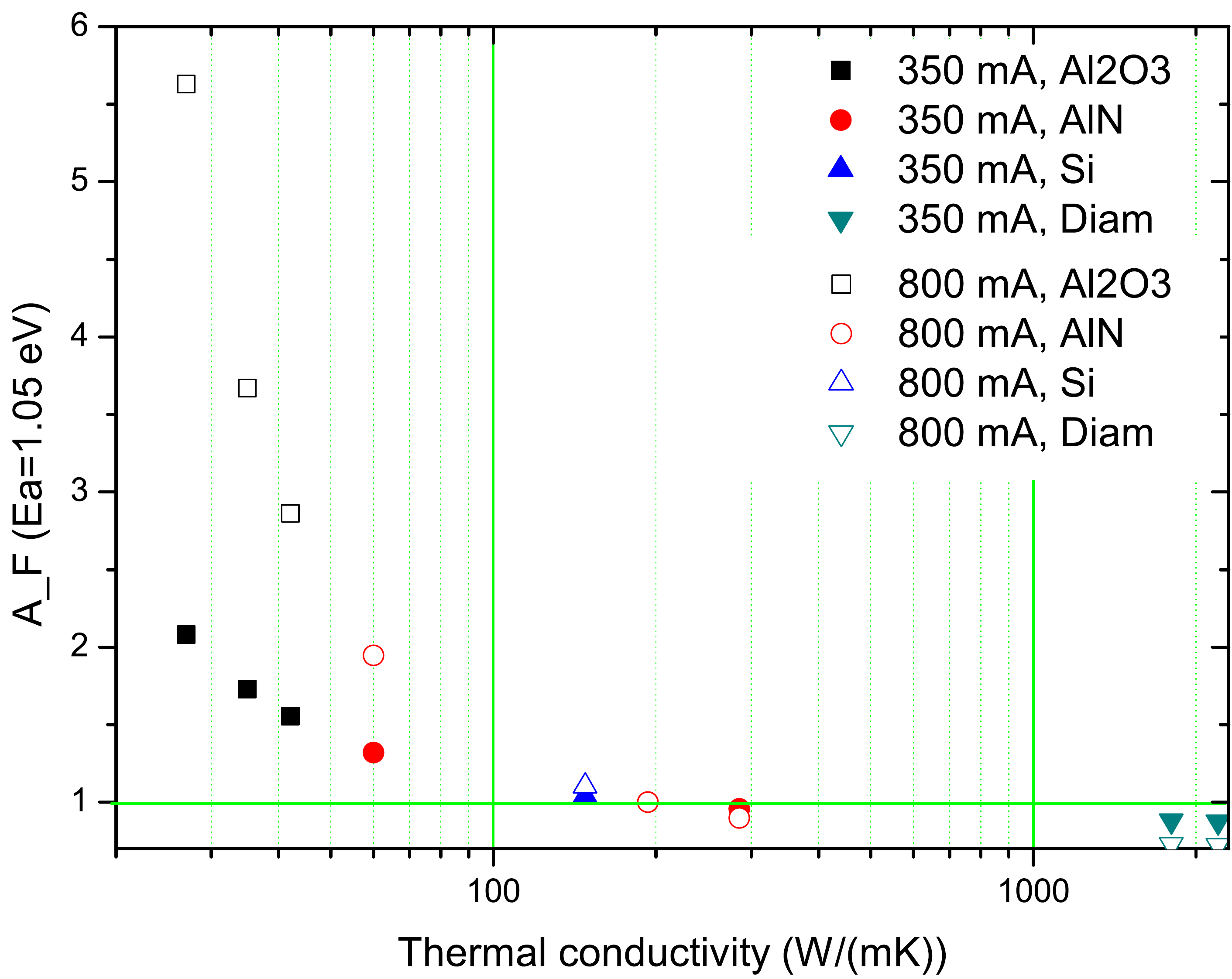}
		\caption{}
		\label{fig:AF_1.05eV}
	\end{subfigure}
	\qquad
	
	\caption{Acceleration factor induced by replacing the AlN (193~W/(m$\cdot$K)) carrier with the other carriers for current levels of 350 and 800~mA. (\subref{fig:AF_2.5eV}) $E_{\rm A}$=2.5~eV activation energy (logarithmic Y scale). (\subref{fig:AF_1.05eV}) $E_{\rm A}$=1.05~eV (linear Y scale).}\label{fig:AF}
\end{figure}

In the current case, we can use the AF to estimate the increase/decrease in the lifetime when a given carrier is replaced with another one with higher/lower $\kappa$. For a given $I$, if one considers the $T_{\rm J}$ obtained with AlN ($\kappa$=193~W/(m$\cdot$K)) as a reference, the AF can be defined as:
\begin{equation}
	AF_{\rm AlN193}=\frac{TTF_{\rm AlN193}}{TTF_{\rm car}}
	 = e^{\frac{q\cdot E_{\rm a}}{k_{\rm B}}\cdot\left(\frac{1}{T_{\rm J,AlN193}}-\frac{1}{T_{\rm J,car}}\right)},
	\label{eq:AF_AlN}
\end{equation}
where $T_{\rm J,AlN193}$ is the junction temperature with the AlN (193~W/(m$\cdot$K)) carrier and $T_{\rm J,car}$ is the junction temperature obtained with the other carriers. The values of $E_{\rm A}$ reported in the literature for GaN devices range between 1.05 and 2.5~eV, reflecting the differences in the processes and materials used by the different laboratories and companies around the world~\cite{Bahl2015}. In the lack of data relative to the Cree LEDs, the AF was calculated for the minimum and maximum values of activation energy found in the literature and for two levels of current, 350~mA and 800~mA. The results are presented in Fig.~\ref{fig:AF}. The impact of the carrier on the lifetime is more evident for higher activation energies (Fig.~\ref{fig:AF_2.5eV}, logarithmic Y scale). For 350~mA, replacing the AlN (193~W/(m$\cdot$K)) with the lowest conductivity Al\textsubscript{2}O\textsubscript{3} carrier will accelerate the aging of the LED by about 6 times. For the same current, replacing the AlN with the diamond carriers reduces the aging of the LED by $\simeq$25\%. If operated at 800~mA the difference becomes more evident: the LED ages 60 times faster with Al\textsubscript{2}O\textsubscript{3} and two times slower with diamond. This means that when operated at this higher current level, the lifetime of a GaN LED with 2.5~eV activation energy mounted on a diamond carrier will double relatively to an LED mounted with an AlN (193~W/(m$\cdot$K)) carrier. Again no significant difference is observed between the performance of SCD and PCD diamond carriers. On the other hand, the crystalline form of AlN increases the lifetime of the LED die by 10\% (350~mA) and $\simeq$20\% (800~mA) with respect to the ceramic AlN with $\kappa$=193~W/m$\cdot$K. For an activation energy of 1.05~eV, the impact of the holders is not as large (Fig.~\ref{fig:AF_1.05eV}, linear scale). The lowest conductivity Al\textsubscript{2}O\textsubscript{3} carrier accelerates the aging of the LED by a factor of $\simeq$2 and $\simeq$6 for 350 and 800~mA, respectively, while the diamond carrier slows down the aging of the LED by $\simeq$10\% and $\simeq$30\% at the same current levels.

\begin{figure}
	\centering
		\centering
		\includegraphics[width=0.35\textwidth]{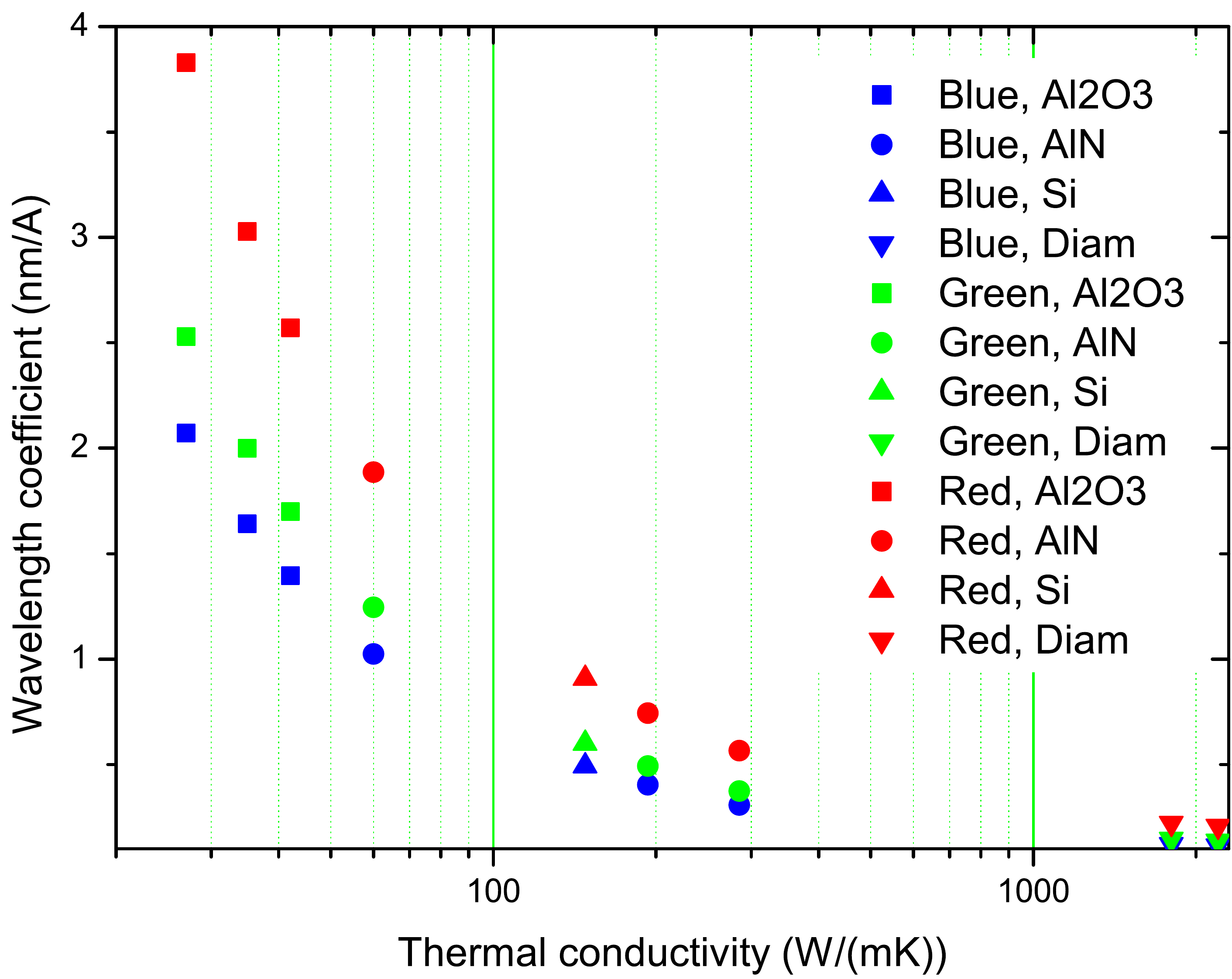}
		\caption{Coefficient of variation of $\lambda$ with $I$.}
		\label{fig:Coeff_lambda_current}
\end{figure}

The final parameter that is directly influenced by $T_{\rm J}$ is the wavelength of the emitted radiation, $\lambda$. When the thermal energy of the charge carriers $k_{\rm B}\cdot T$ ($k_{\rm B}$ being the Boltzmann constant and $T$ the temperature) is small compared to the bandgap energy $E_{\rm g}$, the frequency of the emitted photons ($\nu$) is given by:
\begin{equation}
	\nu=E_{\rm g}/h
	\label{eq:nu},
\end{equation}
where $h$ is the Planck's constant. On the other hand, the wavelength of the emitted photons can be calculated as $\lambda=c/\nu$, where $c$ is the velocity of light. Combining this with Eq.~(\ref{eq:nu}) gives the following expression:
\begin{equation}
	\lambda=\dfrac{c\cdot h}{E_{\rm g}}
	\label{eq:lambda}.
\end{equation}

The dependency of the bandgap energy $E_{\rm g}$ on the temperature $T$ is commonly expressed by the empirical Varshni equation~\cite{Varshni1967}:
\begin{equation}
	E_{\rm g}(T)=E_{\rm g}(0)-\dfrac{\alpha \cdot T^2}{t+\beta}
	\label{eq:Varshni},
\end{equation}
where $E_{\rm g}(0)$ is the bandgap energy at a temperature of 0~K and $\alpha$ and $\beta$ are empirical parameters characterizing the particular semiconductor material.

To estimate the dependency of the wavelength of the blue, green and red Cree\textregistered\space LEDs on the carrier material, the wavelength at 300~K was initially determined from the device datasheet. The bandgap energy at 300~K ($E_{\rm g}(300)$) was calculated by replacing the wavelength in Eq.~(\ref{eq:lambda}). Considering $\alpha=9.4\times10^{-4}$~eV/K and $\beta=791$~K~\cite{Nepal2005} and replacing $E_{\rm g}(300)$ in Eq.~(\ref{eq:Varshni}), $E_{\rm g}(0)$ was finally determined. The calculated values are listed in Table \ref{table:Varshni_parameters}.

Following the extraction of these parameters, the values of $E_{\rm g}$ for the blue, green and red LEDs and the different carriers/current levels were calculated by replacing the simulated values of $T_{\rm J}$ in Eq.~(\ref{eq:Varshni}). Finally, the values of $E_{\rm g}(T)$ were replaced in Eq.~(\ref{eq:lambda}), allowing the determination of $\lambda$ for all the carriers and current values. It was seen that $\lambda$ increases linearly with $I$. The slope of each $\lambda(I)$ curve was determined from the graphs and the results are plotted in Fig.~\ref{fig:Coeff_lambda_current}. For each carrier, the drift of $\lambda$ with $I$ is higher for the red LED and smaller for the blue LED. These results are a consequence of the larger bandgap of blue LEDs (2.51~eV) in comparison to the red LEDs (1.85~eV). The drift of $\lambda$ decreases with the increase in the $\kappa$ of the carrier: the maximum values of $\simeq$3.8/2.1~nm/A for the red/blue LEDs are obtained with the Al\textsubscript{2}O\textsubscript{3} (27~W/(m$\cdot$K)). When the same LED dies are mounted on the diamond carriers, the $\lambda$ drift decreases to $\simeq$ 0.2 and 0.1~nm/A for the red and blue LEDs, respectively.

\begin{table*}\centering
	\begin{tabular}{cccc}\toprule
		\bf\multirow{2}{*}{LED}   	& $\lambda(300 K)$  	& $E_{\rm g}(300 K)$    & $E_{\rm g}(0 K)$\\
		~		                    & nm                    & eV		 	        & eV\\
		\midrule
		Blue   						& 467    				& 2.51		            & 2.59\\
		Green		            	& 515    				& 2.28     		        & 2.35\\
		Red			 			    & 633    				& 1.85  	   	        & 1.93\\
		\bottomrule
	\end{tabular}
	\caption{Wavelength and bandgap energy for blue, green and red Cree\textregistered\space LEDs.}
	\label{table:Varshni_parameters}
\end{table*}

\subsection{Advantages of diamond carriers}

As expected, all the LEDs characteristics improve when materials with higher $\kappa$ are used as die carriers and the choice of a given material over the others should be made taking into consideration the required performance of the LED. Lower cost  Al\textsubscript{2}O\textsubscript{3} and AlN ceramic holders with different $\kappa$ are available from a variety of manufacturers. Diamond holders can be purchased from a few vendors and are available in two forms. SCD plates with $\kappa$ as high as 2200~W/(m$\cdot$K) are manufactured by high power high temperature (HPHT) method~\cite{Gracio2010}; this material shows the lowest concentration of defects and, consequently, the best properties. However, the area of SCD crystals is limited to a few mm\textsuperscript{2}. On the other hand, larger area PCD wafers deposited by chemical vapour deposition (CVD) show a larger number of defects, nevertheless $\kappa$ remains as high as 1800~W/(m$\cdot$K). Curiously, the simulations showed that the performance of the LED dice mounted SCD of PCD carriers is similar.

The replacement of the currently used AlN carrier with a diamond one would improve the reliability of the Cree\textregistered LEDs at different levels. In terms of light intensity, the impact of the carrier is minimum. The amber LED hows the largest change in the light intensity. When compared to the RLF of a LED mounted on an AlN carrier with $\kappa$=193~W/(m$\cdot$K), replacing it with an AlN carrier with 193~W/(m$\cdot$K) or a diamond carrier results in $\simeq$96.4\% and $\simeq$101.7\% of the intensity of light. It can thus be concluded that the $\kappa$ of the carrier has a minor impact on the intensity of the emission.

The impact of the carrier on the lifetime, is quite relevant. For nominal current level and considering the 193~W/(m$\cdot$K) AlN carrier as a reference, replacing it with diamond increases the lifetime to 10\%/25\% at 350~mA for the lowest activation energy, and 25\%/50\% at 800~mA for the highest activation energy.

The diamond carriers also improve the stability of the $\lambda$ of the emitted light with the temperature; the drift of the $\lambda$ is as low as 0.2~nm/A for the red LED and 0.1~nm/A for the blue and green LEDs. With the AlN (193~W/(m$\cdot$K)) carrier, the drift increases to 0.7, 0.5 and 0.4~nm/A for the same LEDs, respectively.

Finally, the use of diamond carriers minimizes the LED footprint. The temperature at the edge of the AlN (193~W/(m$\cdot$K)) carrier is 40.23 and 40.58\textdegree C for 350 and 800~mA, respectively (Fig.~\ref{fig:Edge_temp}). On the other hand, the temperature at the edge of the diamond carrier remains as low as 40.2\textdegree C for 800~mA (Fig.~\ref{fig:Edge_temp}). Looking at Fig.~\ref{fig:800_TempMap_Diam_2200_CSV} it becomes clear that, even for high current levels, the footprint of the die is considerably smaller than that of the carrier, allowing the use of smaller carriers and the increase of the total power density. 

CVD diamond can be the material of choice for very demanding applications, such as the space industry. As an example, let us consider the Lisa (Laser Interferometer Space Antenna) Pathfinder ESA mission, that intends to test in flight the concept of gravitational wave detection by putting two test masses in a near-perfect gravitational free-fall and controlling and measuring their motion with unprecedented accuracy~\cite{ESA}. LISA Pathfinder discharge system currently exploits the photoelectric effect using UV radiation emitted by mercury (Hg) lamps following the method demonstrated by Gravity Probe B~\cite{Buchman2000}. Since the development of this mission, UV LEDs have become commercially available from different manufacturers. These LEDs offer many advantages over traditional Hg lamps, such as a lower mass and volume, increased electrical efficiency, faster response times and the possibility of using light of a shorter wavelength~\cite{Hollington2017}. Even though the results reported in this paper were obtained with Cree\textregistered\ white and different coloured power LEDs, the results can be directly extrapolated for UV LEDs. The heat management of UV LED dies would be dramatically improved by mounting the LED die directly on a diamond carrier, with a positive impact on the lifetime, on the overall emission efficiency, on the stability of the emitted UV radiation, and on the footprint of the devices. Altogether this means that with the diamond carrier the current levels that can be injected in the LEDs without compromising the lifetime would be maximized, and, at the same time, the footprint of the LED (and the corresponding volume and weight) would be decreased.

\section{Conclusions}
The impact of the thermal conductivity of the GaN-SiC die carrier on the lifetime, relative luminous flux and wavelength stability of a Cree\textregistered\space power LED was evaluated through thermal simulations performed with Ansys. Different materials were considered, such as ceramic and single crystalline Al\textsubscript{2}O\textsubscript{3} and AlN, crystalline Si, SCD, and PCD. The difference in the junction temperature obtained with Al\textsubscript{2}O\textsubscript{3} (27~W/(m$\cdot$K)) and SCD (2200~W/(m$\cdot$K)) is 7.2\textdegree C and 17.6\textdegree C for a current level of 350 and 800~mA, respectively. The coefficient of variation of the junction temperature with the current for both materials is 24 and 1.4\textdegree C/A, respectively. The dependence of the RLF of blue and green LEDs on the carrier material is not relevant, unlike the amber LED that shows a decrease of 10\% and an increase of 2\% when an AlN carrier (193~W/(m$\cdot$K)) is replaced with Al\textsubscript{2}O\textsubscript{3} (27~W/(m$\cdot$K)) and diamond (2200~W/(m$\cdot$K)), respectively. The lifetime of the LEDs varies considerably with the carrier material. Using  the lifetime of the LED mounted on the AlN (193~W/(m$\cdot$K)) carrier as a reference and an activation energy of 2.5~eV, the AF increases with the Al\textsubscript{2}O\textsubscript{3} (27~W/(m$\cdot$K)) holder by almost 6 times and by more than 60 times for a current level of 350 and 800~mA, respectively, whereas for the same current levels it decreases to 0.75 and less than 0.5 with the diamond (2200~W/(m$\cdot$K)) carrier. Finally, the impact of both materials on the drift of the wavelength with the current level was also evaluated, being minimum for the blue LEDs and maximum for the red LEDs. For the later LEDs the drift  was 3.8 and 0.2~nm/A with Al\textsubscript{2}O\textsubscript{3} (27~W/(m$\cdot$K)) and diamond (2200~W/(m$\cdot$K)), respectively.
The quality of Al\textsubscript{2}O\textsubscript{3} and AlN carriers has a considerable impact on all the figures, unlike what happens with both SCD and PCD carriers, that have comparable thermal performance. 
Given the tremendous impact of the carrier material on the junction temperature and consequently on all the LED figures, the choice of the carrier material is highly dependent on the required performance of the LED. For applications such as the gravitational wave detection, the use of PCD carriers may increase the lifetime and improve the performance of the UV LED light sources, which are critical for the success of the mission.

\section*{Acknowledgment}
The authors would like to thank Cree\textregistered\space for providing information regarding the structure of the LED and ANSYS support team for providing technical support during the simulations.
This work is funded by FCT/MCTES through national funds and when applicable co-funded by EU funds under the project UIDB/50008/2020-UIDP/50008/2020. Dr. Joana C. Mendes was hired by Instituto de Telecomunicações under the decree law Nr. 57/2016.



\end{document}